\documentclass[twoside,12pt]{article}  

\usepackage{indentfirst}  
\usepackage{bm}    
\usepackage{graphicx}  

\begin{document}    

\begin{center}
\LARGE\bf Fidelity susceptibility and geometric phase in critical phenomenon$^{*}$   
\end{center}

\footnotetext{\hspace*{-.45cm}\footnotesize $^*$Project supported by the National Natural Science Foundation of China (Grant No 11075101), Shanghai Leading
Academic Discipline Project (Project No S30105), Shanghai Research
Foundation (Grant No 07d222020), and the Training Fund of NENU'S
Scientific Innovation Project under Grant No NENU-STC08018.}
\footnotetext{\hspace*{-.45cm}\footnotesize $^\dag$E-mail:
tianlijun@staff.shu.edu.cn }

\begin{center}
\rm Tian Li-Jun$^{\rm a)b)\dagger}$, \ \ Zhu Chang-Qing$^{\rm a)b)}$, \ \ Zhang Hong-Biao$^{\rm c)}$, \ and \ Qin Li-Guo$^{\rm a)b)}$
\end{center}

\begin{center}
\begin{footnotesize} \sl
${}^{\rm a)}$Department of Physics, Shanghai University, Shanghai 200444, China \\   
${}^{\rm b)}$Shanghai Key Lab for Astrophysics, Shanghai 200234, China \\   
${}^{\rm c)}$Institute of Theoretical Physics, Northeast Normal University, Changchun 130024, China \\    
\end{footnotesize}
\end{center}

\vspace*{2mm}

\begin{center}
\begin{minipage}{15.5cm}
\parindent 20pt\footnotesize
Motivated by recent development in quantum fidelity and fidelity
susceptibility, we study relations among Lie algebra, fidelity
susceptibility and quantum phase transition for two-state system and
the Lipkin-Meshkov-Glick model. We get the fidelity susceptibility
for $SU(2)$ and $SU(1,1)$ algebraic structure models. From this
relation, the validity of the fidelity susceptibility to signal for
the quantum phase transition is also verified in these two systems.
At the same time, we obtain the geometric phase in these two systems
in the process of calculating the fidelity susceptibility. In
addition, the new method of calculating fidelity susceptibility has
been applied to explore the two-dimensional XXZ model and the Bose-Einstein
condensate(BEC).
\end{minipage}
\end{center}

\begin{center}
\begin{minipage}{15.5cm}
\begin{minipage}[t]{2.3cm}{}\end{minipage}
\begin{minipage}[t]{13.1cm}

\end{minipage}\par\vglue8pt
{ PACS: 03.65.Fd, 03.67.-a, 03.65.Vf}
\end{minipage}
\end{center}

\section{Introduction}

Fidelity, one of the most intriguing feature of quantum
information science,$^{[1]}$ has been widely studied
in recent years.$^{[2-14]}$
The quantum phase transition(QPT), driven purely by quantum fluctuations and occurring at zero
temperature, is believed to be an important concept in condensed
physics.$^{[15]}$ In a quantum many-body system, the QPT is
driven by purely the quantum fluctuation in ground states. The
wave function of the ground state can have a abrupt change as the system
varies across the phase transition point. Therefore, an approach to
quantum phase transitions based on the quantum-information concept
of fidelity has been put forward.$^{[2]}$ However, except for
a few specific models,$^{[2,16]}$ the calculation of the
ground-state fidelity is tedious. Recently, a neater and simpler
formalism of fidelity to critical phenomena was
introduced,$^{[17]}$ for the so-called fidelity
susceptibility, to signal the whole QPT. The main advantage of this
approach lies in the fact that the fidelity is a purely
Hilbert-space geometrical quantity and no a priori knowledge of the
structure of the considered system is required for its use. As shown
in Ref. [17], the fidelity susceptibility is intrinsically
related to the dynamic structure factor of the driving Hamiltonian
that is evaluated though the scheme based on some
numerical techniques including exact diagonalization and density
matrix renormalization group. On the other hand Zhang \emph{et al}.$^{[18]}$ employed Lie algebra to evaluate
the fidelity susceptibility, and show high efficiency of it.

In this paper, we investigate, in a general framework, how the differential
form of the  fidelity susceptibility can be established in terms of
the general Lie algebra. As will be seen below, under certain conditions
, the differential form is  an effective tool in
detecting the critical points of the QPT. To demonstrate this, analytic formulas for the
fidelity and the fidelity susceptibility are derived for the $SU(2)$ and $SU(1,1)$.
By using these formulas, the fidelity and the fidelity
susceptibility can be easily calculated for a large class of
many-body systems, as long as the Hamiltonian of the system can be
rewritten as the form [see Eq. (4)]. Employing our general formulas
to the two-state system, we can show the fidelity
susceptibility of the system in terms of $SU(2)$. At the same time, the geometric
phase, which is also an effective indicator in detecting the QPT,$^{[19-22]}$ can be obtained in the process of calculating the fidelity
susceptibility. On the other hand, according to the general
expressions of the differential form of the fidelity susceptibility, one can
also expect that the same results for the Lipkin-Meshkov-Glick(LMG)
model in terms of $SU(1,1)$. Furthermore, we extend this differential form
fidelity susceptibility to other physics models.

\section{Formulism}
The general Hamiltonian of quantum many-body systems reads
\begin{equation}
H(\lambda)=H_{0}+\lambda H_{I},
\end{equation}
where $H_{I}$ is the driving Hamiltonian and $\lambda$ denotes its
strength. The fidelity is the absolute value of the overlap between
two ground states $|\Psi_0(\lambda)\rangle$ and
$|\Psi_0(\lambda+\delta\lambda)\rangle$,$^{[2]}$
\begin{equation}
F(\lambda,\lambda+\delta\lambda)=|\langle\Psi_0(\lambda)|\Psi_0(\lambda+\delta\lambda)\rangle|
\end{equation}
with $\delta\lambda$ a small deviation. Extracting the leading term of the fidelity, the fidelity
susceptibility can be obtained$^{[17]}$
\begin{equation}
\chi_{F}(\lambda)=\sum_{n\neq0}\frac{|\langle\Psi_{n}(\lambda)|H_{I}|\Psi_{0}(\lambda)|^{2}}{[E_{n}(\lambda)-E_{0}(\lambda)]^{2}}
\end{equation}
with eigenvalues $E_{n}(\lambda)$ and corresponding normalized
eigenvectors $|\Psi_{n}(\lambda)\rangle$. The eigenstates define a
set of orthogonal complete bases in the Hilbert space.

 Given a physical system, the Hamiltonian can be written as
\begin{equation}
\label{standard}
H=\sum\limits_{i}\epsilon_{i}H_{i}+\sum\limits_{\alpha}(\lambda_{\alpha}E_{\alpha}
+\lambda^{*}_{\alpha}E_{-\alpha}),
\end{equation}
where $\epsilon_{i}$ and $\lambda_{\alpha}$ are coupling parameters
and $\{H_{i},E_{\alpha},E_{-\alpha}\}$ are the Cartan-Weyl basis of
a semi-simple Lie algebra. Using the unitary operator
$\hat{U}(\xi_{\alpha}(\lambda))=\exp[\sum\limits_{\alpha}(\xi_{\alpha}E_{\alpha}-\xi^{*}_{\alpha}E_{-\alpha})]$,
 the
Hamiltonian can be turned into the diagonal form
\begin{equation}
\label{dia}
\hat{U}^{\dag}(\xi_{\alpha})H\hat{U}(\xi_{\alpha})=\sum\limits_{i}\eta_{i}H_{i}.
\end{equation}
Therefore, the eigenstates of Hamiltonian (\ref{standard}) are $|\Psi\rangle=\hat{U}(\xi_{\alpha})|ref\rangle$, where
$|{ref}\rangle$ is the direct product of normalized eigenstates of
$H_i$. Then the absolute value of the overlap between
$|\Psi(\lambda)\rangle=\hat{U(\xi_{\alpha}(\lambda))}|ref\rangle$
and
$|\Psi(\lambda+\delta\lambda)\rangle=\hat{U}(\xi_{\alpha}(\lambda+\delta\lambda))|ref\rangle$
is
\begin{equation}
F=|\langle{ref}|\hat{U}^{\dag}(\xi_{\alpha}(\lambda)\hat{U}(\xi_{\alpha}(\lambda+\delta\lambda)))|ref\rangle|.
\end{equation}
According to the general relation between fidelity and fidelity
susceptibility,
$F=1-\frac{1}{2}(\delta\lambda)^{2}\chi(\lambda)+\cdots$,
the expression of the fidelity susceptibility is given by$^{[18]}$
\begin{equation}
\label{new}
\chi(\lambda)=-\langle{ref}|(\hat{U}^{\dag}\partial_{\lambda}\hat{U})^2|{ref}\rangle-|\langle{ref}|\hat{U}^{\dag}\partial_{\lambda}\hat{U}|{ref}\rangle|^2.
\end{equation}

\section{Two-state systems}
The two-state system is the simplest quantum system which can be calculated exactly. Furthermore, the two-state
systems possess several typical quantum properties. So we shall study it firstly. The
Hamiltonian of a two-state system can be written as
\begin{equation}
\label{tss} H=-{\textbf{B}}\cdot\mathbf{\sigma},
\end{equation}
where ${\textbf{B}}$ is an external magnetic field and
$\sigma=(\sigma_{x}, \sigma_{y}, \sigma_{z})$ are Pauli matrices. In
$\sigma_{z}$ basis $|\uparrow\rangle, |\downarrow\rangle$, Pauli
matrices take the form
\begin{eqnarray}
\sigma_{x}=  \left(
\begin{array}{cc}
0 & 1\\
1 & 0
\end{array}
\right),  \sigma_{y}=  \left(
\begin{array}{cc}
0 & -i\\
i & 0
\end{array}
\right), \sigma_{z}=  \left(
\begin{array}{cc}
1 & 0\\
0 & -1
\end{array}
\right).
\end{eqnarray}
We can rewrite $H$ in the $SU(2)$ form
\begin{equation}
H=-2B\cos\theta J_{z}-B\sin\theta e^{i\phi}J_{+}-B\sin\theta
e^{-i\phi}J_{-},
\end{equation}
by means of the generators of the algebra $SU(2)$
\begin{equation}
J_{z}=\frac{1}{2}\sigma_{z}, J_{+}=\frac{1}{2}\sigma_{+},
J_{-}=\frac{1}{2}\sigma{-}.
\end{equation}
These satisfy the usual commutation relations
\begin{equation}
[J_{+},J_{-}]=2J_{z}, [J_{z},J_{\pm}]=\pm{J_{\pm}}.
\end{equation}
Resorting to the Eq. (\ref{dia}), we introduce the unitary operator
$\hat{U}(\theta,
\phi)=\exp(-\frac{\theta}{2}e^{-i\phi}J_{+}+\frac{\theta}{2}e^{i\phi}J_{-})$.
The Hamiltonian (\ref{tss}) then can be diagonalized
\begin{equation}
\hat{U}^{\dag}(\theta, \phi)H\hat{U}(\theta, \phi)=-BJ_{z}.
\end{equation}
Here, $\theta$ and $\phi$ can be regarded as adiabatic
parameters. For simplicity and without loss of generality, we fixed
$\theta$ first and have
\begin{equation}
\hat{U}^{\dag}(\theta,
\phi)\partial_{\phi}\hat{U}({\theta,\phi})=\frac{i}{2}\sin\theta
(e^{-i\phi}J_{+}+ e^{i\phi}J_{-})+2i\sin^{2}\frac{\theta}{2}J_{z}
\end{equation}
and
\begin{equation}
 |\langle ref|\hat{U}^{\dag}(\theta, \phi)\partial_{\phi}\hat{U}(\theta,
\phi)|ref\rangle|^{2}=4\sin^{2}\frac{\theta}{4},
\end{equation}
where $|ref\rangle$ is the $\sigma_{z}$
basis $|\uparrow\rangle, |\downarrow\rangle$ in the two-state system. Finally we get the
fidelity susceptibility of spin-$\frac{1}{2}$ subjected to an external
magnetic field
\begin{equation}
\label{fs}
 \chi_{F}=\frac{1}{4}\sin^{2}\theta.
\end{equation}

Since $\langle ref|\hat{U}^{\dag}(\theta,
\phi)\partial_{\phi}\hat{U}(\theta, \phi)|ref\rangle$ is just the
Berry adiabatic connection, which contribute a Pancharatnam-Berry
phase$^{[23,24]}$ to the spin as the magnetic field rotates
adiabatically around cone direction, we get the Berry
phase
\begin{eqnarray}
\label{berry}
 \gamma(\theta, \phi)&&=-i\int^{2\pi}_{0}\langle
ref|\hat{U}^{\dag}(\theta,
\phi)\partial_{\phi}\hat{U}(\theta, \phi)|ref\rangle\nonumber\\
&& =\pm\pi(1-\cos\theta).
\end{eqnarray}
The results of Eqs. (\ref{fs}) and (\ref{berry}) are in agreement
with that in Ref. [25], which verify the reliability of our method.

\section{The Lipkin-Meshkov-Glick model}
The LMG model$^{[26-28]}$ was originally introduced in
nuclear physics. It provides a
simple description of the tunneling of bosons between two degenerate
levels and can thus be used to describe many physical systems, such
as two-mode Bose-Einstein condensates$^{[29]}$ and Josephson
junctions.$^{[30]}$ Recently the entanglement in this
model has attracted much interest because of available numerical
calculations and plentiful phase diagrams.$^{[31]}$ In the
thermodynamic limit, its phase diagram can be simply established by
a semiclassical approach.$^{[32]}$ For finite large number $N$ of particles, it was studied by the $1/N$ expansion in the
Holstein-Primakoff single boson representation$^{[33]}$ and by
the continuous unitary transformation.$^{[34]}$

The Hamiltonian of the LMG model can be written as
\begin{eqnarray}
\label{lmg}
H&&=-\frac{\lambda}{N}\sum_{i<j}(\sigma^{i}_{x}\sigma^{j}_{x}+\gamma\sigma^{i}_{y}\sigma^{j}_{y}+\gamma\sigma^{i}_{y}\sigma^{j}_{y})-h\sum_{i}\sigma^{i}_{z}\nonumber\\
&&=-\frac{2\lambda}{N}(S^{2}_{x}+\gamma
S^{2}_{y})-2hS_{z}+\frac{\lambda}{2}(1+\gamma)\nonumber\\
&&=-\frac{\lambda}{N}(1+\gamma)({\textbf{S}}^{2}-S^{2}_{z}-\frac{N}{2})-2hS_{z}\nonumber\\
&&-\frac{\lambda}{2N}(1-\gamma)(S^{2}_{+}+S^{2}_{-}),
\end{eqnarray}
where the $\sigma_{\kappa}(\kappa=x,y,z)$ are the Pauli matrices,
$S_{\kappa}=\sum_{i}{\sigma^{i}_{\kappa}}/{2}$, and
$S_{\pm}=S_{x}\pm iS_{y}$. $\lambda$ and $h$ are the spin-spin
interaction strength and the effective external field, respectively.
$N$ is the total spin numbers and ${1}/{N}$ ensures that the
free energy per spin is finite in the thermodynamical limit. It is
understood that $H$ preserve the magnitude of the total spin and the
parity $P=\prod_{i}\sigma^{i}_{z}$, i.e.,
\begin{equation}
[H, {\textbf{S}}^{2}]=0,   [H, P]=0,
\end{equation}
for all values of the anisotropy parameter $\gamma$. Specially, in
the isotropic case $\gamma=1$, one has $[H, S_{z}]=0$, so
that $H$ is diagonal in the eigenbasis of ${\textbf{S}}^{2}$ and
$S_{z}$. We adopt the ${1}/{N}$ expansion method corresponding to
the large $N$ limit. We first use the Holstein-Primakoff boson
representation of the spin operator$^{[33]}$ in the
$S={N}/{2}$ subspace given by
\begin{eqnarray}
&&S_{z}=S-a^{\dag}a=\frac{N}{2}-a^{\dag}a,\nonumber\\
&&S_{+}=(2S-a^{\dag}a)^{\frac{1}{2}}a=N^{\frac{1}{2}}(1-\frac{a^{\dag}a}{N})^{\frac{1}{2}}a=S^{\dag}_{-},
\end{eqnarray}
where the standard bosonic creation and annihilation operator
satisfy $[a, a^{\dag}]=1$. This representation is well adapted to
the computation of the low-energy physics with $\langle a^{\dag}a
\rangle/N\ll1$.

The next step consists in inserting these expression in Eq.
(\ref{lmg}), and to expand the argument of the square roots. Keeping
terms of order $(1/N)^{-1}$, $(1/N)^{-1/2}$, and $(1/N)^{0}$ in the
Hamiltonian yields ($h\geq1$)
\begin{equation}
\label{su}
H=-hN+(2h-1-\gamma)a^{\dag}a-[(1-\gamma)/2](a^{\dag2}+a^{2}).
\end{equation}

In order to analyze the fidelity susceptibility of the above
equation, we transform Eq. (\ref{su}) into
the form of $SU(1,1)$ following the ideas in Ref. [35].
We first introduce the generators of $SU(1,1)$,
\begin{equation}
K_{+}=\frac{1}{2}a^{\dag2}, K_{-}=\frac{1}{2}a^{2},
K_{z}=\frac{1}{4}(2a^{\dag}a+1),
\end{equation}
 which satisfy the  communication relations of $SU(1,1)$ algebra. Submitting these expressions of the $SU(1,1)$ generators into
Eq. (\ref{su}), one can get
\begin{equation}
H=-hN-\frac{1}{2}(h-\gamma-1)+2(2h-\gamma-1)K_{z}+(\gamma-1)K_{+}+(\gamma-1)K_{-},
\end{equation}
which consists with Eq.
(\ref{standard}). We do not show the diagonalization of $SU(1,1)$
algebraic structure model explicitly here, but interested readers
are recommended to refer to Ref. [18]. We simply quote the
main result that is connection with the fidelity susceptibility of
the $SU(1,1)$ algebra model. After all the procedures, the fidelity
susceptibility of $SU(1,1)$ algebraic structure model becomes
\begin{equation}
\label{su1,1}
\chi_{SU(1,1)}=\frac{1}{8}[(\frac{\partial\theta}{\partial\lambda})^{2}+\sinh^{2}\theta(\frac{\partial\phi}{\partial\lambda})].
\end{equation}
Assuming $\tanh\theta=(1-\gamma)/(2h-1-\gamma)$, the fidelity
susceptibility is represented by
\begin{equation}
\chi_{F}=\frac{(1-\gamma)^{2}}{32(1-h)^{2}(h-\gamma)^{2}},
\end{equation}
which is in agreement with Ref. [36].

\vspace*{4mm}


\begin{center}
\begin{figure}[h]
\includegraphics[angle=0,width=7cm]{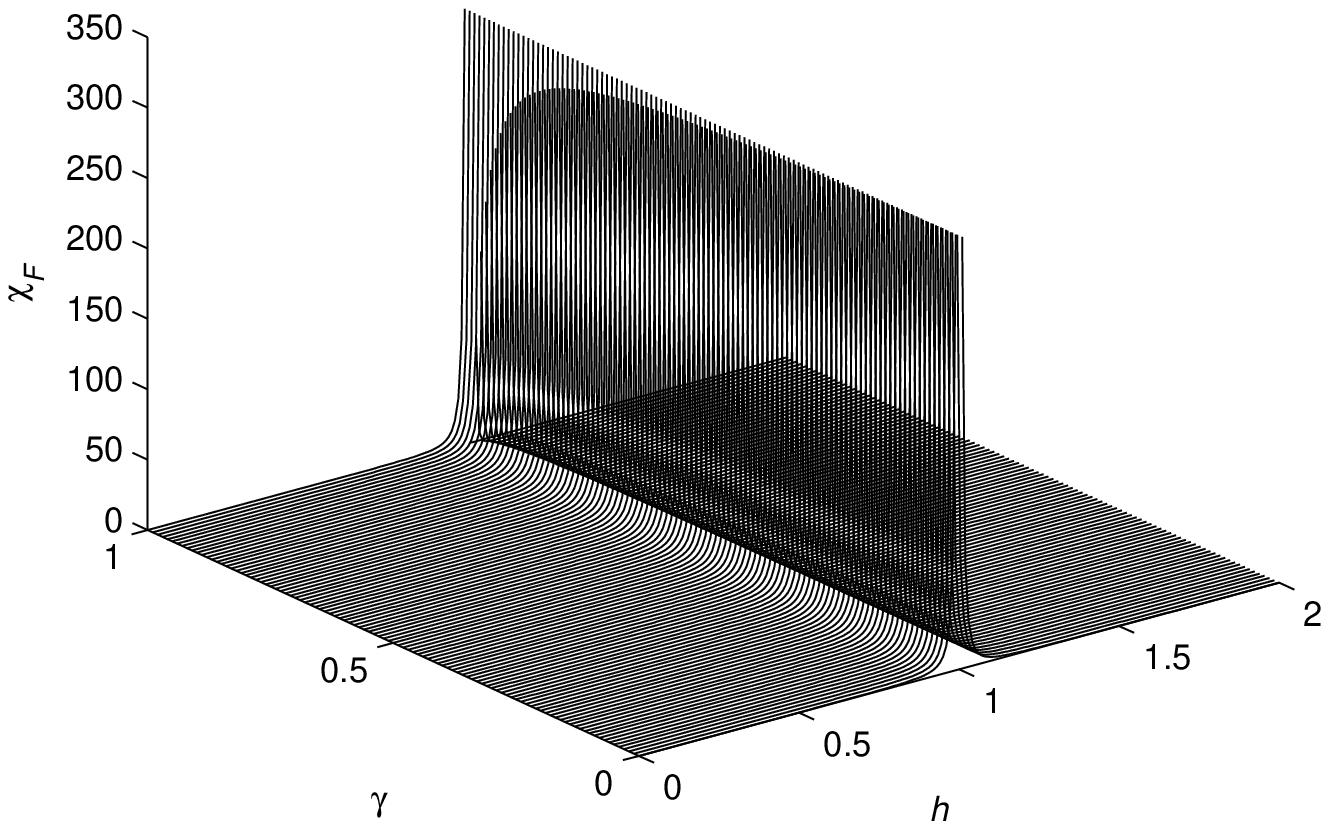}\\
\parbox{15.5cm}{\small{\bf Fig.1.} Fidelity susceptibility as a function of $h$ and $\gamma$.
The divergent character of $\chi_{F}$ is clearly displayed as
$h\rightarrow1$.}
\end{figure}
\end{center}

\vspace*{4mm}


\begin{center}
\begin{figure}[h]
\includegraphics[angle=0,width=7cm]{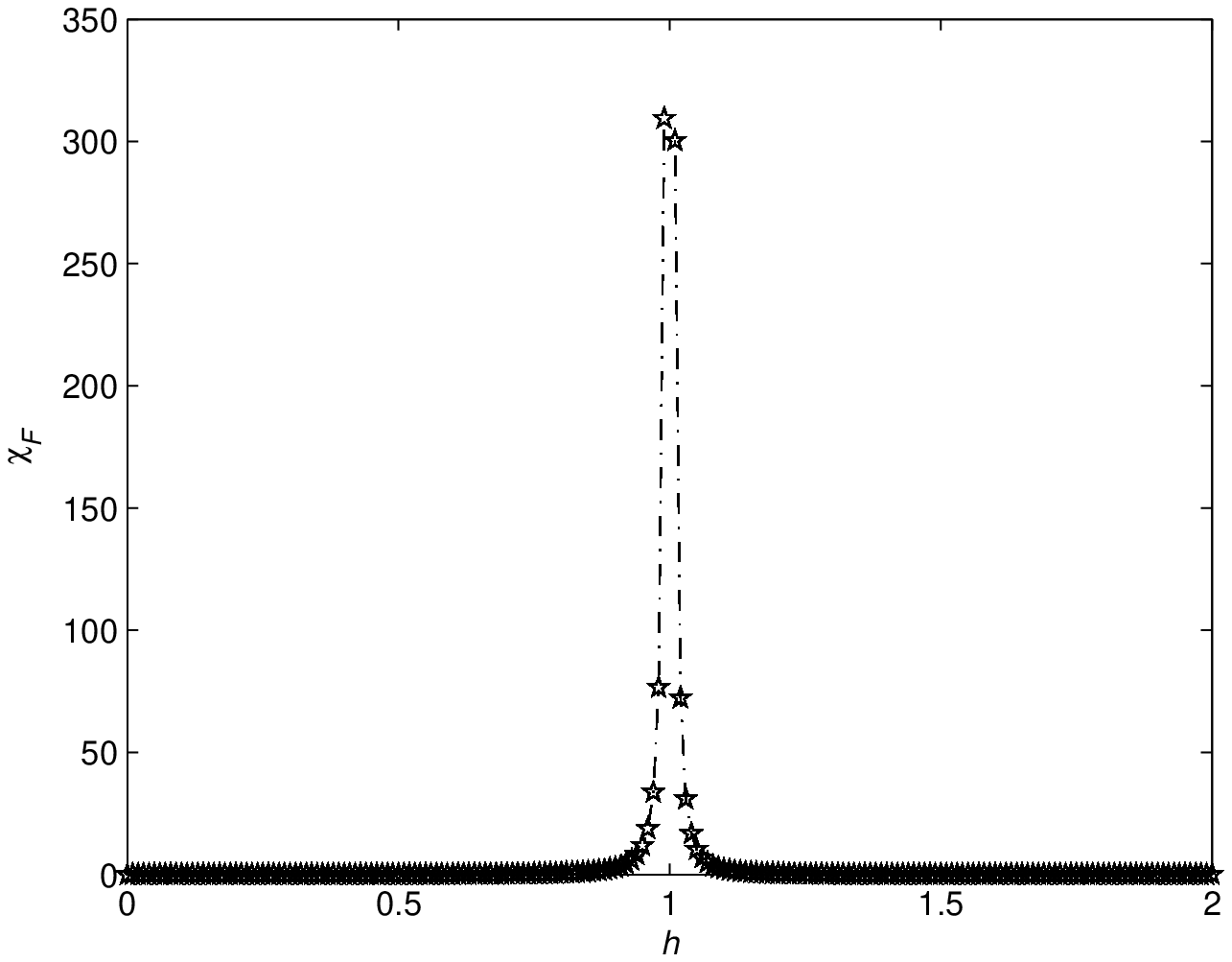}\\
\parbox{15.5cm}{\small{\bf Fig.2.} Fidelity susceptibility as a function of $h$ for
$\gamma=0.5$ in large N limit. }
\end{figure}
\end{center}

The derivation above is only valid for $h\geq1$, for $0<h<1$ the
calculation is actually similar to the above case of $h\geq1$. When
$0<h<1$, $\tanh\theta=\frac{h^{2}-\gamma}{2-h^{2}-\gamma}$, the
fidelity susceptibility is then obtained accordingly
\begin{equation}
\chi_{F}=\frac{h^{2}}{8(1-h^{2})^{2}}.
\end{equation}
Thus we obtained fidelity susceptibility of the anisotropic LMG
model in the large $N$ limit. As shown in Figs. 1 and 2, it is obvious that
$\chi_{F}$ is divergent at $h=1$ where the LMG model has
been proved to experience a second-order phase transition
independent  of the anisotropy $\gamma$, which is well described by
a mean-field approach.$^{[34]}$
 Through the calculation of fidelity susceptibility, we have found the geometric phase. For illustration, we consider the system which has a
rotation $g(\phi)$ around the new $z$ direction. The Hamiltonian
becomes $H(\phi)=g(\phi)Hg^{\dag}(\phi)$ with $g(\phi)=e^{i\phi
S_{z}}$. Then Eq. (\ref{su}) can be written as
\begin{equation}
H=-hN+(2h-1-\gamma)a^{\dag}a-[(1-\gamma)/2](a^{\dag2}e^{-2i\phi}+a^{2}e^{2i\phi}).
\end{equation}
The geometric phase of the ground state, accumulated by varying the
angle $\phi$ from 0 to $\pi$, is described by
$\beta=-i\int_{0}^{\pi}\langle g| \frac{\partial}{\partial\phi}
|g\rangle$.$^{[19]}$ Finally, we get the
geometric phase
\begin{eqnarray}
\beta&&=-i\int_{0}^{\pi}\langle g| \frac{\partial}{\partial\phi}
|g\rangle=-i\int_{0}^{\pi}\langle1,0|U(\theta,
\phi)\partial_{\phi}U(\theta,\phi)|1,0\rangle\nonumber\\
&&=-i\int_{0}^{\pi}-2i\sinh^{2}\frac{\theta}{2}=\pi(1-\cosh\theta).
\end{eqnarray}
The derivation above is also only valid for $h\geq1$.
When $0<h<1$, $\tanh\theta=\frac{h^{2}-\gamma}{2-h^{2}-\gamma}$.
Therefore we can get geometric phase of the LMG model in the
thermodynamic limit. One can find $\beta_{g}$ is also divergent at $h=1$ which
is equal to the fidelity susceptibility in indicating the quantum
phase transition in Figs. 3 and 4.
\vspace*{4mm}


\begin{center}
\begin{figure}[h]
\includegraphics[angle=0,width=7cm]{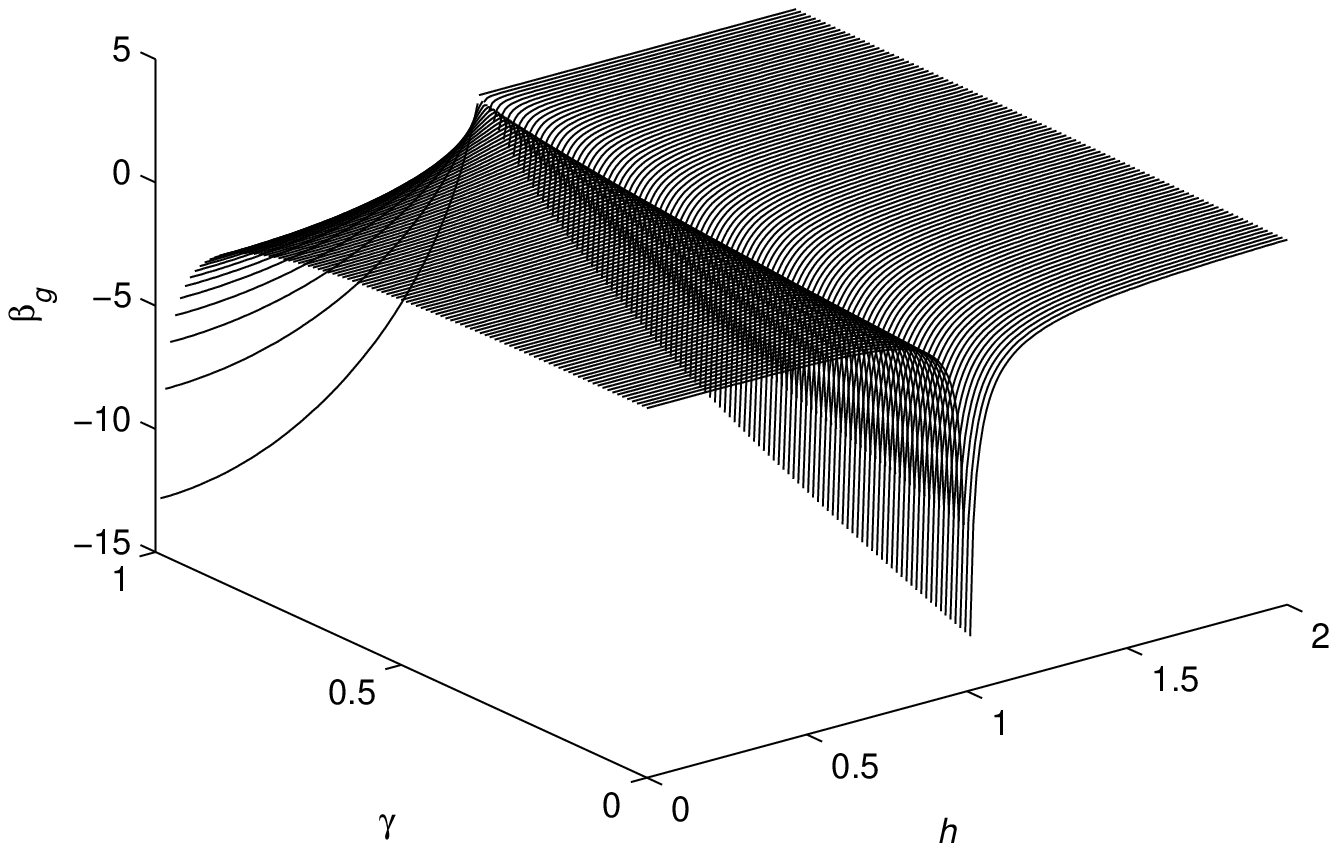}\\
\parbox{15.5cm}{\small{\bf Fig.3.} The geometric phase as a function of $h$ and $\gamma$. It
is obvious that $\beta_{g}$ is divergent at $h=1$
independent of $\gamma$.}
\end{figure}
\end{center}

\vspace*{4mm}


\begin{center}
\begin{figure}[h]
\includegraphics[angle=0,width=7cm]{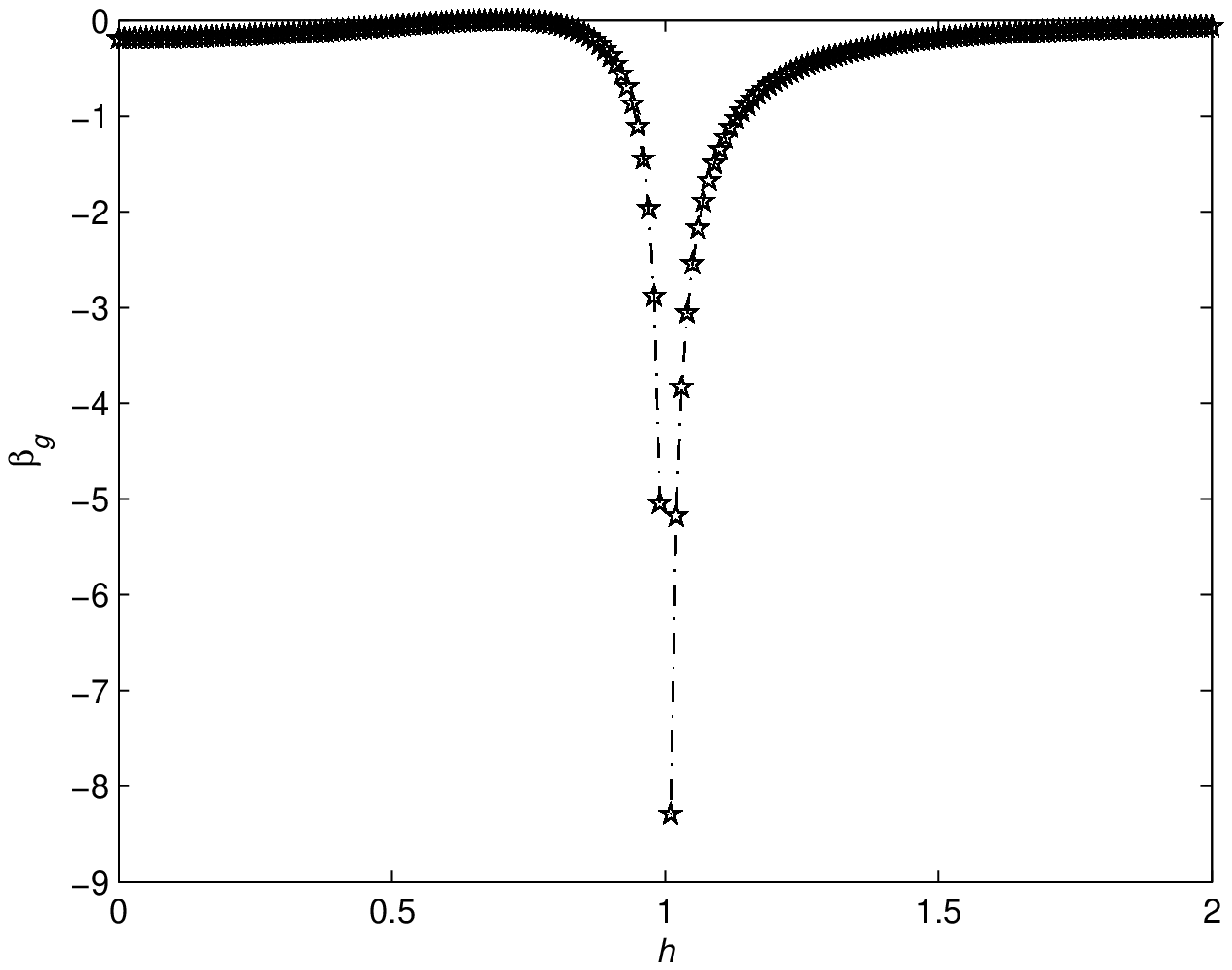}\\
\parbox{15.5cm}{\small{\bf Fig.4.} The geometric phase as a function of $h$. The parameter
$\gamma=0.5$.}
\end{figure}
\end{center}

\section{Other models}
In this section, we generalize the new method of calculating
fidelity susceptibility to other quantum many-body models. As there
is nobody to obtain the analytical results of the fidelity susceptibility of the
two-dimensional XXZ model and the Bose-Einstein condensate precisely, we
expect the under results can arouse more wonderful ideas or results.
\subsection{Two-dimensional XXZ model}
The Hamiltonian of the $XXZ$ antiferromagnetic model reads
\begin{equation}
H_{XXZ}=\sum_{\langle{ij}\rangle}(S^{x}_{\bf{i}}S^{x}_{\bf{j}}+S^{y}_{\bf{i}}S^{y}_{\bf{j}}+\eta
S^{z}_{\bf{i}}S^{z}_{\bf{j}}),
\end{equation}
where $S^{\alpha}_{i}(\alpha=x,y,z)$ are the spin-$1/2$ operators at
site $i$ and $\eta=J_{z}/J_{x}(J_{x}=J_{y})$ is a dimensionless
parameter characterizing the anisotropy of the model. The sum runs
over all the nearest neighbors on a square lattice. We begin with
the two-sublattice model and Holstein-Primakoff
transformation,$^{[33]}$
\begin{eqnarray}
\label{x5}
&&S^{+}_{a}=\sqrt{2S}a^{\dag}(1-\frac{a^{\dag}a}{2S})^{1/2},S^{-}_{a}=(S^{+}_{a})^{\dag},\nonumber\\
&&S^{+}_{b}=\sqrt{2S}b^{\dag}(1-\frac{b^{\dag}b}{2S})^{1/2},S^{-}_{b}=(S^{+}_{b})^{\dag},\nonumber\\
&&\hat{S}^{z}_{a}=-S+a^{\dag}a, \hat{S}^{z}_{b}=S-b^{\dag}b,
\end{eqnarray}
where $a^{\dag}$, $a (b^{\dag}, b)$ are boson creation and
annihilation operators on sublattice A (sublattice B), respectively.
The particle numbers $n_{a}=a^{\dag}a$, $n_{b}=b^{\dag}b$ cannot
excel $2S$. Transforming the operators into momentum space, we
rewrite the Hamiltonian as
\begin{equation}
H_{XXZ}=-2\eta zNS^{2}+2zS\sum_{\bf{k}}H_{\bf{k}},
\end{equation}
where $z$ is the coordination number of the lattice and $H_{k}$ is of the form
\begin{equation}
H_{k}=\eta(a^{\dag}_{\bf{k}}a_{\bf{k}}+b^{\dag}_{\bf{k}}b_{\bf{k}})+\gamma_{\bf{k}}(a_{\bf{k}}b_{\bf{k}}+a^{\dag}_{\bf{k}}b^{\dag}_{\bf{k}}).
\end{equation}
Here
$\gamma_{k}=z^{-1}\sum\limits_{\bf{R}}e^{i{\bf{k}}\cdot{\bf{R}}}$
with $\bf{R}$ is a vector connecting an atom with its nearest
neighbor. We can rewrite $H_{\bf{k}}$ in the $SU(1,1)$ form,
\begin{equation}
H_{k}=2\eta{A^{\bf{k}}_{z}}+\gamma_{k}(A^{\bf{k}}_{+}+A^{\bf{k}}_{-}),
\end{equation}
by means of the generators of the algebra $SU(1,1)$
\begin{eqnarray}
&&A^{\bf{k}}_{+}=a^{\dag}_{\bf{k}}b^{\dag}_{\bf{k}},
A^{\bf{k}}_{-}=a_{\bf{k}}b_{\bf{k}},\nonumber\\
&&A^{\bf{k}}_{z}=\frac{1}{2}(n^{a}_{\bf{k}}+n^{b}_{\bf{k}}+1).
\end{eqnarray}
Taking a similar transformation of $H_{\bf{k}}$, we get the
unitary operator
$U=\exp(\xi_{\bf{k}}A^{\bf{k}}_{+}-\xi_{\bf{k}}A^{\bf{k}}_{-})$ with
$\tan\xi_{\bf{k}}=-\frac{\gamma_{k}}{\eta}$. With the spin-wave
theory framework, we obtain the unitary operator and then calculate the
fidelity susceptibility of the model in two dimensions. Taking
$\tanh\theta=\frac{\gamma_{k}}{\eta},
e^{-i\Phi}=\frac{\gamma_{k}}{\gamma_{k}}=1$ into the Eq.
(\ref{su1,1}), one can obtain
\begin{equation}
\chi=\int\frac{(\eta\partial_{\lambda}\gamma_{k}-\gamma_{k}\partial_{\lambda}\eta)^{2}}{32(\eta^{2}-\gamma^{2}_{k})^{2}}\frac{d\bf{k}}{(2\pi)^{3}}.
\end{equation}
\subsection{Fidelity susceptibility for the Bose-Einstein condensate}
The standard description of the Bose-Einstein condensate is via an
order parameter field $\Psi(x)$. The Hamiltonian takes the standard
form
\begin{eqnarray}
\label{x10} \mathcal
{H}[\Psi]=\int{d^{3}x[\frac{\hbar^2}{2m}|\nabla\Psi(x)|^{2}+U(x)|\Psi(x)|^2]}\nonumber\\
+\frac{1}{2}\int{d^{3}x}\int{d^{3}y\Psi^{*}(y)\Psi^{*}(x)V(x,y)\Psi(y)\Psi(x)}.
\end{eqnarray}
As in Ref. [35], the Eq. (\ref{x10}) can be rewritten in the
form of $SU(1,1)$ algebra
\begin{eqnarray}
\mathcal
{H}=2[\sigma_{0}A^{0}_{3}+\frac{1}{2}(u_{0}A^{0}_{+}+u^{*}_{0}A^{0}_{-})]\nonumber\\
+\sum\limits_{k\neq0}[\sigma_{k}A^{k}_{3}+\frac{1}{2}(u_{k}A^{k}_{+}+u^{*}_{k}A^{k}_{-})]-E_{*},
\end{eqnarray}
where
$\sigma_{0}\equiv\epsilon_{0}+\frac{1}{2}\sum\limits_{k\neq0}(V_{0}+V_{k})(\langle{n_{k}}\rangle+\langle{n_{-k}}\rangle)$,
$u_{0}\equiv
V_{0}\langle{a_{0}^{2}}\rangle+\sum\limits_{k\neq0}V_{k}\langle{a_{k}a_{-k}}\rangle$,
$\sigma_{k}\equiv\epsilon_{k}+\langle{n_{0}}\rangle(V_{0}+V_{k})$,
$u_{k}\equiv{V_{k}}\langle{a^{2}_{0}}\rangle$,
$E_{*}=\frac{1}{2}[V_{0}|\langle{a^{2}_{0}}\rangle|^{2}+\sigma_{0}]+\frac{1}{2}\sum\limits_{k\neq0}[(\sigma_{k}-\epsilon_{k})\langle{n_{k}+n_{-k}}\rangle+\sigma_{k}]+\frac{1}{2}\sum\limits_{k\neq0}(u_{k}\langle{a^{+}_{k}a^{+}_{-k}}\rangle+u^{*}_{k}\langle{a_{k}a_{-k}}\rangle)$.
Introducing the generators of the algebra $SU(1,1)$
\begin{equation}
A^{0}_{3}=\frac{1}{2}(n_{0}+\frac{1}{2}),
A^{0}_{+}=\frac{a^{+2}_{0}}{2}, A^{0}_{-}=\frac{a^{2}_{0}}{2}
\end{equation}
and
\begin{equation}
A^{k}_{3}=\frac{1}{2}(n_{k}+n_{-k}+1),A^{k}_{+}=a^{+}_{k}a^{+}_{-k},A^{k}_{-}=a_{k}a_{-k}.
\end{equation}
We calculate the fidelity susceptibility of $\mathcal
{H}_{0}=2[\sigma_{0}A^{0}_{3}+\frac{1}{2}(u_{0}A^{0}_{+}+u^{*}_{0}A^{0}_{-})]$.
From this equation, we can get
\begin{equation}
U=\exp(\xi_{0}A^{0}_{+}-\xi^{*}_{0}A^{0}_{-}), \xi_{0}=r\exp(i\phi).
\end{equation}
Taking the same procedures as in the two-dimensional XXZ model, we can get the
fidelity susceptibility of the $\mathcal {H}_{0}$
\begin{eqnarray}
\chi_{0}=\frac{[\sigma_{0}(u_{0}\partial_{\lambda}u^{*}_{0}+u^{*}_{0}\partial_{\lambda}\sigma)-2|u_{0}|^{2}\partial_{\lambda}\sigma_{0}]^{2}}{32|u_{0}|^{2}(\sigma^{2}_{0}-|u_{0}|^{2})^{2}}\nonumber\\
-\frac{(\sigma^{2}_{0}-|u_{0}|^{2})(u^{*}_{0}\partial_{\lambda}u_{0}-u_{0}\partial_{\lambda}u^{*}_{0})^{2}}{32|u_{0}|^{2}(\sigma^{2}_{0}-|u_{0}|^{2})^{2}}.
\end{eqnarray}

Similarly the fidelity susceptibility of the $\mathcal {H}_{k}$ can
be obtained
\begin{eqnarray}
\chi_{k}=\frac{[\sigma_{k}(u_{k}\partial_{\lambda}u^{*}_{k}+u^{*}_{k}\partial_{\lambda}\sigma)-2|u_{k}|^{2}\partial_{\lambda}\sigma_{k}]^{2}}{32|u_{k}|^{2}(\sigma^{2}_{k}-|u_{k}|^{2})^{2}}\nonumber\\
-\frac{(\sigma^{2}_{k}-|u_{k}|^{2})(u^{*}_{k}\partial_{\lambda}u_{k}-u_{k}\partial_{\lambda}u^{*}_{k})^{2}}{32|u_{k}|^{2}(\sigma^{2}_{k}-|u_{k}|^{2})^{2}}.
\end{eqnarray}

\section{Conclusions}
In conclusion, we have established the differential form of the
fidelity susceptibility in terms of the general Lie algebras.
Meanwhile we  investigate the geometric phase which can also show the
phase transition point. Therefore, we construct the
relation between the fidelity susceptibility and geometric phase. We also apply the differential form of fidelity
susceptibility to other physics models. In
particular, we focus on the $SU(2)$ and $SU(1,1)$ algebras. The form
of the fidelity susceptibility of $SU(2)$ and $SU(1,1)$ algebra not
only enables us to evaluate the fidelity susceptibility easily, but
also builds a straightforward connection between quantum-information
theory and the Lie algebra in quantum many-body physics.


\end{document}